\documentclass{jfm}
\usepackage{graphicx}
\usepackage{epstopdf, epsfig}
\usepackage[export]{adjustbox}
\usepackage[amssymb]{SIunits}
\usepackage{xcolor} 
\usepackage{amsfonts}
\usepackage{latexsym}
\usepackage{amssymb}
\usepackage[tight,footnotesize]{subfigure}
\usepackage{booktabs}
\usepackage[utf8]{inputenc}
\usepackage{longtable}
\usepackage{array}
\usepackage{tikz}
\usepackage{wasysym}
\usepackage[normalem]{ulem}
\usepackage[section]{placeins}
 \mathchardef\mhyphen="2D

\definecolor{height_max}{rgb}{0.6350, 0.0780, 0.1840}
\definecolor{height_mid}{rgb}{0.4660, 0.6740, 0.1880}
\definecolor{height_min}{rgb}{0.0, 0.45, 0.73}

\definecolor{deepcarrotorange}{rgb}{0.91, 0.41, 0.17}
\definecolor{mtlb_orange}{rgb}{0.9290, 0.6940, 0.1250}
\definecolor{mtlb_blue}{rgb}{0.0, 0.45, 0.73}
\definecolor{mtlb_green}{rgb}{0.4660, 0.6740, 0.1880}
\definecolor{mtlb_red}{rgb}{0.6350, 0.0780, 0.1840}
\definecolor{mtlb_purple}{rgb}{ 0.4940, 0.1840, 0.5560}

\shorttitle{Heat transport in homogeneous bubbly flow}
\shortauthor{B. Gvozdi\'{c} et al.}

\title{Experimental investigation of heat transport in homogeneous bubbly flow}

\author{Biljana Gvozdi\'{c}\aff{1}, Elise Alm\'{e}ras\aff{1,2}, Varghese Mathai\aff{1}, Xiaojue Zhu\aff{1}, Dennis P. M. van Gils\aff{1}, Roberto Verzicco\aff{1,3}, Sander G. Huisman\aff{1}, Chao Sun\aff{4,1}\corresp{\email{chaosun@tsinghua.edu.cn}}, \and Detlef Lohse\aff{1}}

\affiliation{\aff{1}Physics of Fluids Group, J. M. Burgers Center for Fluid Dynamics and Max Planck Center Twente, Faculty of Science and Technology, University of Twente, P.O. Box 217, 7500 AE Enschede, The Netherlands
\aff{2}Laboratoire de G\'{e}nie Chimique, UMR 5503, CNRS-INP-UPS, 31106 Toulouse, France
\aff{3}Department of Mechanical Engineering, University of Rome `Tor Vergata', Rome 00133, Italy
\aff{4}Center for Combustion Energy and Department of Thermal Engineering, Tsinghua University, 100084 Beijing, China}

\begin{document}

\maketitle

\begin{abstract}

We present results on the global and local characterisation of heat transport in homogeneous bubbly flow. 
Experimental measurements were performed with and without the injection of $\sim \unit{2.5}{\milli \meter}$ diameter bubbles (corresponding to $Re_b \approx 600$) in a rectangular water column heated from one side and cooled from the other. 
The gas volume fraction~$\alpha$ was varied in the range  $0\% - 5\%$, and the Rayleigh number $Ra_H$ in the range $4.0 \times 10^9 - 1.2 \times 10^{11}$.
We find that the global heat transfer is enhanced up to 20 times due to bubble injection.
Interestingly, for bubbly flow, for our lowest concentration $\alpha = 0.5\% $ onwards, the Nusselt number $\overline{Nu}$ is nearly independent of $Ra_H$, and depends solely on the gas volume fraction~$\alpha$. 
We observe the scaling $\overline{Nu} \propto \alpha^{0.45}$, which is suggestive of a diffusive transport mechanism, as found by \cite{almeras2015mixing}.  
Through local temperature measurements, we show that the bubbles induce a huge increase in the strength of liquid temperature fluctuations, e.g. by a factor of 200 for $\alpha = 0.9\%$.  
Further, we compare the power spectra of the temperature fluctuations for the single- and two-phase cases. 
In the single-phase cases, most of the spectral power of the temperature fluctuations is concentrated in the large scale rolls/motions. 
However, with the injection of bubbles, we observe intense fluctuations over a wide range of scales, extending up to very high frequencies. 
Thus, while in the single-phase flow the thermal boundary layers control the heat transport, once the bubbles are injected, the bubble-induced liquid agitation governs the process from a very small bubble concentration onwards. 
Our findings demonstrate that the mixing induced by high Reynolds number bubbles ($Re_b \approx 600$) offers a powerful mechanism for heat transport enhancement in natural convection systems.

\end{abstract}


\section{Introduction}

Enhancing the heat transport in flows is desirable in many practical applications.
To achieve this in systems with natural convection, several approaches have been adopted.
For example, in vertical natural convection, the usage of fins or riblets is proven to increase the heat flux \citep{shakerin1988natural}. 
In Rayleigh-B\'{e}nard convection \citep{ahlers2009heat,lohse2010small} heat transfer enhancement can be achieved by tuning the boundary conditions to aid the formation of large scale rolls or coherent structures \citep{chong2015condensation}, or by introducing wall roughness elements (e.g. \cite{roche2001observation,tisserand2011comparison,xie2017turbulent,zhu2017roughness}).  
While these methods have been widely used to optimize convective transport, they pose limits on the maximum achievable heat flux in moderate to high Rayleigh number flows.

An alternative is the injection of bubbles in the flow, which indeed enhances the heat transport considerably \citep{deckwer1980mechanism}. 
In general, the bubbles can be injected in a quiescent liquid phase (``pseudo-turbulent" flow \citep{risso2002velocity,riboux2010experimental,roghair2011energy,mercado2010bubble}) or in an already turbulent liquid phase (turbulent bubbly flow \citep{rensen2005effect,van2006turbulent,van2013importance,spandan2016drag,prakash2016energy}).
It is known that the motion of injected bubbles induces mixing of warm and cold parcels of the liquid phase, which in industrial applications where heat transport is coupled with the bubbly flow can lead to a 100 times greater heat transfer coefficient when compared to the single-phase case \citep{deckwer1980mechanism}.
Therefore there is a practical benefit from a better understanding of the heat transport in bubbly flows as this enables better design and optimization of the industrial processes.
As a result, the effect of bubbles on heat transfer has been subject of several experimental and numerical studies in the past.

The bubbles can be injected in a system with natural or forced convection.
Early studies which focused on forced convective heat transfer in bubbly flows \citep{sekoguch1980forced, sato1981momentum, sato1981momentum2} showed that the bubbles modify the temperature profile and that higher void fractions close to the heated wall lead to an enhanced heat transfer. 
In a more recent numerical study on forced convective heat transfer in turbulent bubbly flow in vertical channels, \cite{dabiri2015heat} showed that both, nearly spherical and deformable bubbles, improve the heat transfer rate. 
They found that a 3\% volume fraction of bubbles increases the Nusselt number by 60\%.

Studies on natural convection in bubbly flow have been performed mostly by introducing micro-bubbles \citep{kitagawa2008heat,kitagawa2009effects} and sub-millimeter-bubbles \citep{kitagawa2013natural} close to the heated wall. 
Among these, the micro-bubbles (mean bubble diameter $d_{bub} = \unit{0.04}{\milli \meter}$) showed higher heat transfer enhancement as compared to sub-millimeter-bubbles ($d_{bub} = \unit{0.5}{\milli \meter}$) \citep{kitagawa2013natural}. 
The authors stated that this occurred because the micro-bubbles form large bubble swarms which rise close to the wall and enhance mixing in the direction of the temperature gradient, while sub-millimeter bubbles, owing to their weak wake and low bubble number density, resulted in limited mixing.
In contrast, in case of bubbles with diameter of a few millimeters, the wake of individual bubbles and vortex shedding behind the bubbles play a significant role in the heat transfer enhancement \citep{kitagawa2013natural}. 
\cite{tokuhiro1994natural} experimentally studied the effects of $\unit{2-4}{\milli \meter}$ diameter inhomogeneously injected nitrogen-bubbles on laminar and turbulent natural convection heat transfer from a vertical heated plate in mercury.
They reported a twofold increase in the heat transfer coefficient as compared to the case without bubbles. 
Similarly, \cite{deen2013direct} showed in their numerical study that a few high Reynolds number bubbles rising in quiescent liquid could increase the local heat transfer between the liquid and a hot wall. 

In systems with natural convection bubbles can be introduced through boiling as well. 
Some of the studies on this subject have been performed for Rayleigh-B\'{e}nard convection.
The Rayleigh-B\'{e}nard system consists of a flow confined between two horizontal parallel plates, where the bottom plate is heated and the top one is cooled  \citep{ahlers2009heat,lohse2010small}.
In case of boiling it has been found that bubbles strongly affect velocity and temperature fields  depending on the Jakob number which is defined as a ratio of latent heat to sensible heat \citep{oresta2009heat,zhong2009enhanced,schmidt2011modification}.
Numerical studies performed by \cite{lakkaraju2013heat} found that without taking into consideration the bubble nucleation and bubble detachment, depending on the number of the bubbles and superheat the heat transfer can be enhanced up to around 6 times.
In an attempt to control the bubble nucleation process, \cite{guzman2016heat,guzman2016vapour} performed an experimental study where they varied the geometry of the nucleation sites of the bubbles and found that the heat transfer could be enhanced up to 50\%.

To summarize, previous studies (performed in systems with natural convection) mostly focused either on heat transfer in inhomogeneous bubbly flow, where bubbles of different sizes were introduced close to the hot wall, and where the thermal stability of the used setups remained unclear, or the studies were performed in a well-defined Rayleigh-B\'{e}nard system, but in which the bubbles mainly consisted of vapor and where bubble volume fraction and the bubble size varied due to evaporation and condensation.

In this study on heat transfer in bubbly flow we choose a different approach.
Firstly, we use a rectangular water column heated from one side and cooled from the other~(see figure~\ref{fig1}) which resembles the classical vertical natural convection system.
Secondly, at the bottom of the setup we homogeneously inject millimetric-bubbles, so that in the bubbly case pseudo-turbulent flow is present, the dynamics of which is adequately characterized and broadly studied in the past.
We characterise the global heat transfer in both the single- and two-phase flow cases with the goal of understanding how the imposed temperature difference (characterised by the the Rayleigh number $Ra_H$) influences the heat flux (characterised by the Nusselt number $\overline{Nu}$) for various gas volume fractions $\alpha$, in order to try to better understand the mechanism of heat transport enhancement.
The characterization of the global heat transfer is based on the calculation of the dimensionless temperature difference, the Rayleigh number:
\begin{equation}
  Ra_H =  \frac{g  \beta  (\overline{T_h}-\overline{T_c})  H^3}{\nu   \kappa};
\end{equation}
and the dimensionless heat transfer rate, the Nusselt number:
\begin{equation}
 \overline{Nu} = \frac{Q / A}{\chi \ (\overline{T_h} - \overline{T_c})/ L  }.
 \end{equation}
Here, $Q$ is the measured power supplied to the heaters, $\overline{T_h}$ and $\overline{T_c}$ are the mean temperatures of the hot and cold walls, respectively, $L$ is the length of the setup, $A$ is the surface of the sidewall, $\beta$ is the thermal expansion coefficient, $g$ the gravitational acceleration, $\kappa$ the thermal diffusivity, and $\chi$ the thermal conductivity of water. 
In this study we choose the height to be the characteristic length scale for $Ra_H$ since the boundary layer regime is present (see Section \ref{global}), where the velocity is predominately in the vertical direction. 
Note that in this study the Prandtl number is nearly constant, $Pr \equiv \frac{\nu}{\kappa} = 6.5 \pm 0.3$.   

Furthermore, we perform temperature profile measurements with and without bubble injection, by traversing a small thermistor along the length of the setup at the mid-height.
In this way we obtain information on the statistics of the temperature fluctuations and on the power spectrum of the temperature fluctuations in a homogeneous bubbly flow.

\section{Experimental setup and instrumentation} \label{experiments}

\subsection{Experimental setup}

\begin{figure}
\centering
\includegraphics[scale=1]{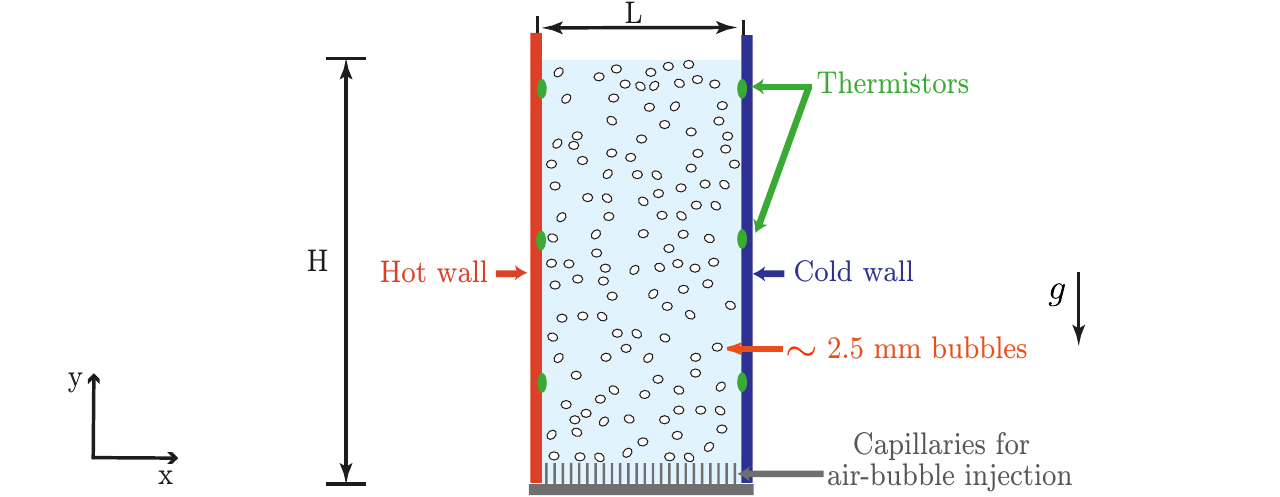}
\caption{Rectangular bubbly column heated from one sidewall and cooled from the other ($H = \unit{600}{\milli \meter}$, $L = \unit{230}{\milli \meter}$). Bubbles of about $ \unit{2.5}{\milli \meter}$ diameter were injected through 180 capillaries placed at the bottom of the setup (inner diameter $\unit{0.21}{\milli \meter}$). 
}
\label{fig1}
\end{figure}

The experiments were performed in a rectangular bubble column ($ 600 \times 230 \times \unit{60}{\milli \meter ^3}$), shown in figure~\ref{fig1}.
Air bubbles of about $ \unit{2.5}{\milli \meter}$ diameter were injected into quiescent demineralized water using 180 capillaries (inner diameter $\unit{0.21}{\milli \meter}$) uniformly distributed over the bottom of the column. 
The gas volume fraction was varied from 0.5\% to 5\% by controlling the inlet gas flow rate via a digital mass flow controller (Bronkhorst F-111AC-50K-AAD-22-V).
The two main sidewalls of the setup ($ 600 \times \unit{230}{\milli \meter ^2}$) were made of $\unit{1}{\centi \meter}$ thick glass and two (heated resp. cooled) sidewalls ($ 600 \times  \unit{60}{\milli \meter ^2}$) of $\unit{1.3}{\centi \meter}$ thick brass. 

As mentioned previously, our setup resembles one for vertical natural convection since one brass sidewall is heated and the other is cooled in order to generate a horizontal temperature gradient in the setup.
More precisely, heating was provided by placing three etched-foil heaters on the outer side of the hot wall.
Heaters were connected in parallel to a digitally controlled power supply (Keysight N8741A), providing altogether up to $\unit{300}{\watt}$.
The other brass wall was cooled by a water circulating bath (Polytemp PD15R-30).
The temperature of the heated and the cooled walls was monitored by three thermistors which were glued on different heights of the cold and warm walls, namely at $\unit{125}{\milli \meter}$, $\unit{315}{\milli \meter}$, and $\unit{505}{\milli \meter}$.
Temperature regulation of both the cold and warm walls was achieved by PID (Proportional-Integral-Derivative) control so that the mean temperature of the walls was maintained constant over time. 
In order to limit the heat losses, the setup was wrapped in several layers of insulating blanket and foam.
Moreover, an aluminium plate with heaters attached to it was placed on the outer side of the hot wall and maintained at the same temperature as the hot wall to act as a temperature shield. 

Heat losses were estimated to be not more than 7\% by calculating convective heat transport rate from all outer surfaces of the setup with the assumption that these surfaces are at maximum $\unit{25}{\celsius}$.
On the other hand, we measured the power needed to maintain the temperature of the bulk constant ($T_{bulk} = \unit{25}{^\circ \Celsius}$) over 4 hours. 
Power supplied to the heaters in that way is not more that 3\% of the total power needed when running the actual experiments in which the bulk temperature is also 25 degrees.
We therefore expect the actual heat losses to be in the range of 3\% to 7\% which enables us to study precisely the heat transport driven by a horizontal temperature gradient in a bubbly flow.

\subsection{Single phase vertical convection}
\begin{figure}
\begin{center}
\includegraphics[scale=1]{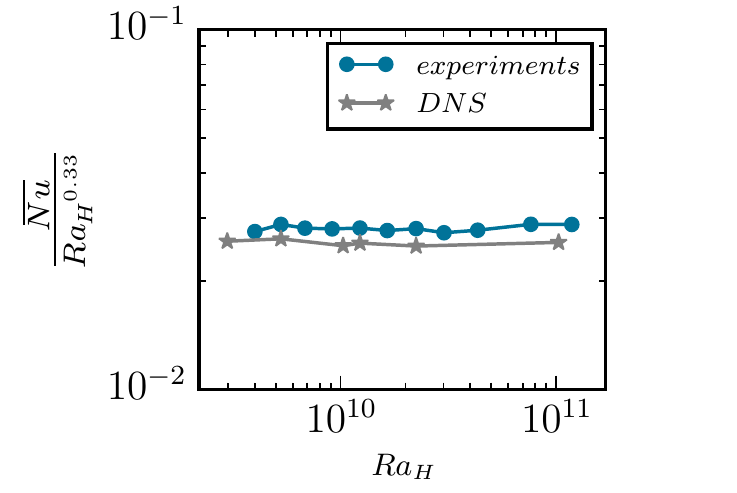}
\end{center}

\caption{Compensated form of the dependence of Nusselt number $\overline{Nu}$ on the Rayleigh number $Ra_H$ for single phase case}
\label{Ra_Nu_sp}
  \end{figure}

The single-phase heat transport will be used as a reference for the heat transport enhancement by bubble injection.  
In a single-phase system a variety of flow regimes can be observed depending on the height $H$, the length $L$, and the Rayleigh number $Ra_H$ of the system \citep{bejan2013convection}.
For the parameter range studied here ($H/L = 2.4$ and $Ra_H = 4.0 \times 10^9 - 1.2 \times 10^{11}$) we expect the system to be in a boundary layer dominant regime. 
In this regime the boundary layers which distinctly form along the heated and cooled sidewalls control the heat transfer, while the bulk of the fluid is relatively stagnant.
Furthermore, previous studies on single-phase vertical convection describe the dependence of Nusselt number on the Rayleigh number in the power law form $Nu\sim Ra^\beta$ at fixed $Pr$, with exponent $\beta$ ranging between $1/4$ and $1/3$ (see \cite{ng2015vertical,ng2017changes} and references within). 
We find that for the range of $Ra_{H}$ studied here the effective scaling exponent is $\beta \approx 0.33 \pm 0.02$, which lies within the expected range as can be seen on the compensated plot in figure \ref{Ra_Nu_sp}. 

For the single phase flow we benchmark the heat transfer against direct numerical simulations (DNS).
For this purpose an in-house second order finite difference code \citep{van2015pencil,zhu17afid} was used to solve the three-dimensional Boussinesq equations for the single phase vertical convection. The code has been extensively validated and used for Rayleigh-B\'enard flow  \citep{van2015pencil,zhu17afid}, in which the only difference is the direction of the buoyancy force. The computational box has the same size as has been employed in the experiments. The no-slip boundary conditions are adopted for the velocity at all solid boundaries. At the top and bottom walls, the heat-insulating conditions are employed, and at the left and right plates, constant temperatures are prescribed. The resolution of the simulations is fine enough to guarantee that the results are grid-independent.
Figure \ref{Ra_Nu_sp} shows good agreement between numerical and experimental results, within the range of uncertainty. 
As the experimental setup was not completely insulated at the top and due to unavoidable heat losses to the outside, the experimentally obtained $\overline{Nu}$  are slightly higher.

\subsection{Instrumentation for the gas phase characterization} \label{gas_char}

\begin{figure} 
\centering
\includegraphics[scale=1]{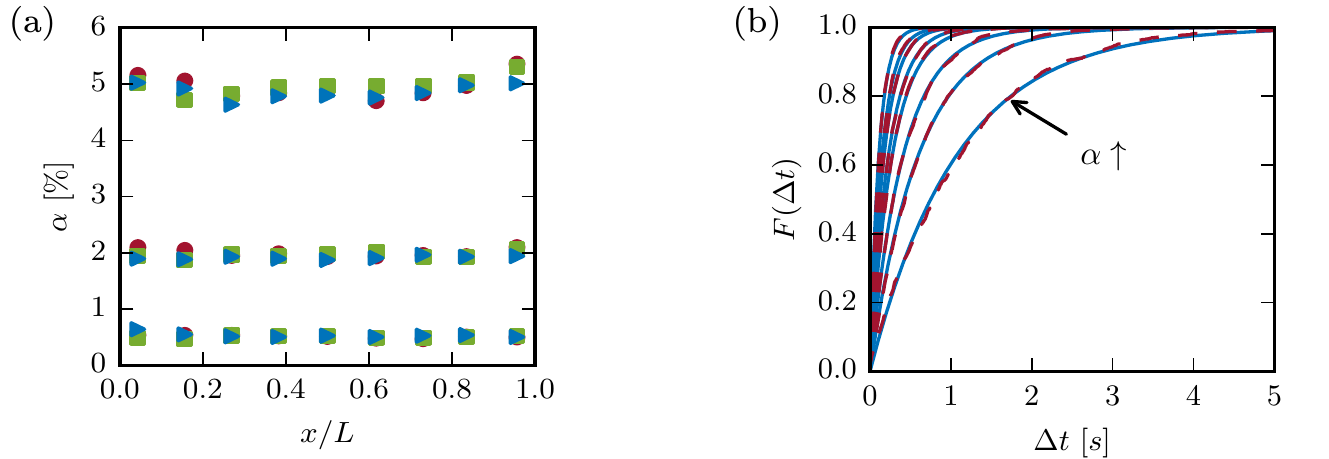}
\caption{(a) Gas volume fraction $\alpha$ at different measurement positions in the experimental setup: $\textcolor{height_min}{\blacktriangleright} \ \mhyphen \ H = \unit{125}{\milli \meter} $, $\textcolor{height_mid}{\blacksquare}\ \mhyphen \ H =\unit{315}{\milli \meter}$, $\textcolor{height_max}{\CIRCLE}\ \mhyphen \ H = \unit{505}{\milli \meter}$; (b) Cumulative distribution function of the time interval between consecutive bubbles in the center of the setup for $ \alpha = 0.3{\%}$,  $0.5{\%}$,  $0.9{\%}$,  $ 1.5{\%}$, $ 2{\%}$, $ 3.0{\%}$, $ 3.9{\%}$ and $ 6.0{\%}$, respectively. }
\label{fig:gvf_clust}
 \end{figure} 
\begin{figure} 
\centering
\includegraphics[scale=1]{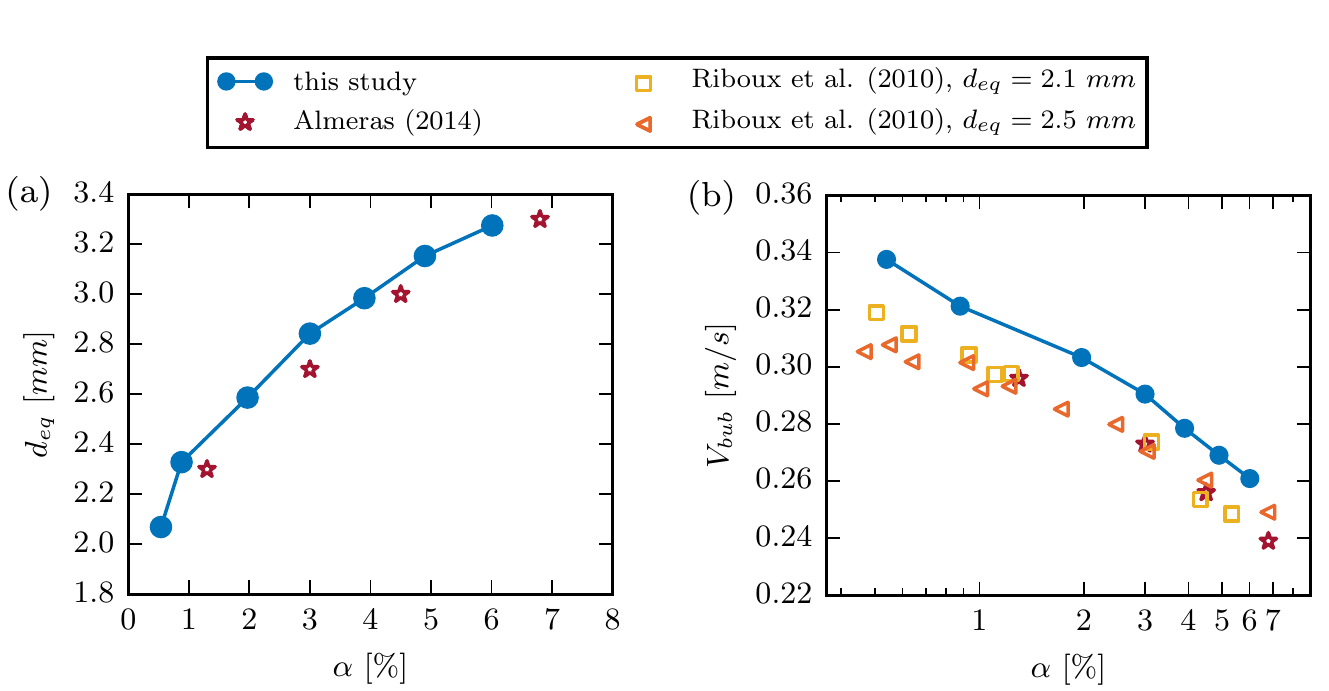}
\caption{(a) Mean bubble diameter $d_{eq}$ and (b) bubble rise velocity $V_{bub}$ for the studied gas volume fractions $\alpha$, in comparison to different experimental studies.}
\label{fig:dia_vel}
 \end{figure} 

The homogeneity of the bubble swarm was verified by measuring the gas volume fraction~$\alpha$ at different locations within the flow by means of a single optical fiber probe. 
In the absence of a temperature gradient, we observe that $\alpha$ is nearly uniform along the length $L$ and height $H$ of the setup~(see figure \ref{fig:gvf_clust} (a)).
We also note that no large-scale clustering is present, since the cumulative distribution function $F(\Delta t)$ of the time between consecutive bubbles $\Delta t$ is Poissonian (see figure \ref{fig:gvf_clust} (b) and \cite{risso2002velocity} for more details).
In presence of the heating, the homogeneity of the swarm was comparable to that without heating, with not more than 2\% variation.
The bubble swarm thus remained homogeneous even with heating, indicating that the bubbly flow was not destabilised by the temperature gradient. 
We further characterised the gas phase by measuring the bubble diameter and the bubble rising velocity with an in-house dual optical fiber probe~\citep{almeras2017experimental}.
Bubble diameter $d_{eq}$ and bubble rise velocity $V_b$ lie in the ranges $[2.1,\ 3.4]\unit{}{\milli \meter}$ and $[0.24,\ 0.34]\unit{} {\meter \per \second}$, respectively~(see figure \ref{fig:dia_vel} (a) \& (b)), resulting in a bubble Reynolds number $Re_b = V_b d_{eq}/\nu \approx 600$.
The bubble rise velocity follows the same trend as previously  observed by \cite{riboux2010experimental}, namely it evolves roughly as $V_{bub}\propto \alpha^{-0.1}$.
These values of bubble diameter (resp. bubble velocities) fall within the expected range for this configuration of needle injection (resp. bubble diameter).
Once compared to those found by \cite{riboux2010experimental} and \cite{almeras2014etude}, we find up to 6\% variation which is acceptable since it can be attributed to slightly different bubble injection section and water quality.

\subsection{Instrumentation for the heat flux and temperature measurements}
In the present study, we performed global and local characterisations of the heat transfer.
In order to obtain $\overline{Nu}$ and  $Ra_H$, we measured the hot and cold wall temperatures, and the heat input to the system $Q$.
To this end, resistances of the thermistors placed on the hot and cold walls were read out every 4.2 seconds using a digital multimeter (Keysight 34970A).
The temperature is then converted from the resistances based on the calibrations of the individual thermistors. 
The total heater power input was measured as  $Q = \sum Q_i = Q_i = I_i \cdot V_i$, where $I_i$ and $V_i$ are the current and voltage across each heater, respectively. 
The experimental measurements were performed after steady state was achieved in which the mean wall temperatures fluctuated less that $ \pm 0.01$ K (resp. $\pm 0.1$ K) for lowest $Ra_H$ and $\pm 0.4$ K (resp. $\pm 0.5$ K) for highest $Ra_H$, for single-phase (resp. two-phase) case.
Time averaging of the instantaneous power supplied to each heater $Q_i$ was then performed over a total time period of 6 hours for single-phase cases, and 3 hours for two-phase cases.

\begin{figure} 
\centering
\includegraphics[scale=1]{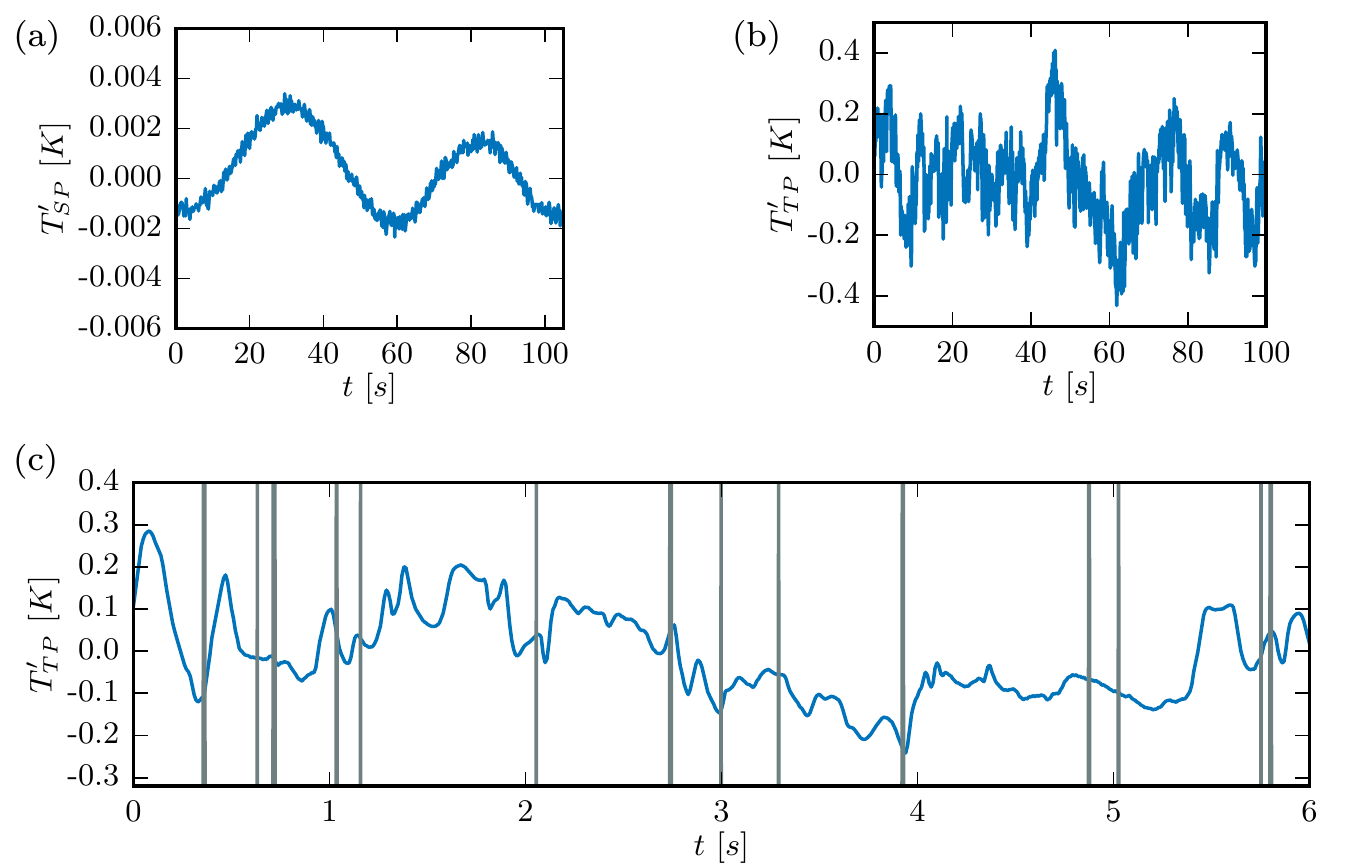}
\caption{Signal of the temperature fluctuations at $ L = \unit{11.5}{\centi \meter}$ for $Ra_H = 2.2 \times 10^{10}$ in (a) single-phase flow $T'_{SP}$ (observed frequency of around $10^{-2}$ Hz corresponds to large scale circulation frequency, which is addressed in Section \ref{local}), (b) two-phase flow $\alpha = 0.9\%$ $T'_{TP}$ , and (c) enlarged simultaneously obtained temperature fluctuations and optical fiber signal where each gray vertical line is a passage of a single bubble.} 
\label{fig:fiber_temp}
 \end{figure}

For a better understanding of the heat transfer, we performed local measurements of the liquid temperature fluctuations: $T'= T - \langle T \rangle$, where $T'$ - temperature fluctuations, $T$ - measured instantaneous temperature and $ \langle T \rangle$ - time averaged temperature at the measurement point. 
For each operating condition, temperature fluctuations were recorded for at least 180 min (resp. 360 min) in two-phase (resp. single-phase) case once the flow was stable, which yields satisfactory statistical convergence.
We used a NTC miniature thermistor manufactured by TE connectivity (Measurement Specialties G22K7MCD419) with a tip diameter of ${\unit{0.38}{\milli \meter}}$, and a response time of ${\unit{30}{\milli \second}}$ in water. 
The thermistor was connected as one arm of a Wheatstone bridge so that very small variations of the thermistors' resistance could be measured. 
For noise reduction we used a Lock-In amplifier (SR830). 
The function of the Lock-In amplifier is to firstly supply voltage to the bridge and then filter out noise at frequencies which are different from that of the signal range. 
This ensured milli-Kelvin resolution of the temperature fluctuations. 
In order to measure the temperature with this resolution reliably, the thermistors were calibrated in a circulating bath with  ${\unit{5}{\milli \kelvin}}$  stability. 
A typical obtained signals in single~-~phase and two-phase are presented in the figure \ref{fig:fiber_temp} a) and b), respectively. 
It is interesting to note that the temperature fluctuations are up to 200 times stronger in the two-phase case and this will be discussed in Section \ref{local}.

The local temperature measurement technique used here is well established for single~-~phase flow \citep{belmonte1994temperature}; however, until now it has never been used for temperature fluctuations measurements in bubbly flows. 
Since the presence of bubbles may perturb the local measurements by interacting directly with the probe \citep{rensen2005effect, mercado2010bubble}, it is  necessary to validate the technique for two-phase flow. 
For this purpose we made an in-house probe in which an optical fiber and the thermistor were positioned $\sim \unit{1}{\milli \meter}$ apart in the horizontal plane and with the thermistor placed $\sim \unit{1}{\milli \meter}$ below the fiber tip.
Figure \ref{fig:fiber_temp} c) shows typical thermistor (blue line) and optical fiber (vertical gray lines) signals simultaneously obtained. 
The optical fiber allows us to detect the presence of the bubble at the probe tip; it was thus possible to remove parts of the temperature signal corresponding to bubble-probe collisions \citep{mercado2010bubble}, and to compare the statistical properties of this truncated temperature signal with the original signal. 
The difference in the statistics (mean, standard deviation, kurtosis, skewness) obtained from the original and the truncated signal was minor ($\approx 0.05\%$).
This suggests that the short durations of the bubble-thermistor contact~($t_{int}=d_{eq}/V_{bub}= \unit{7.2}{\milli \second} $ for $\alpha = 0.9 \% $) were filtered out due to the longer response time of the thermistor~($t_r \sim 30 $ ms). 
Nevertheless, the thermistor response time is sufficiently short to measure the temperature fluctuations between two bubble passages.
From table \ref{table:t_bub} we see that $t_{2b}$ is sufficiently long when compared to $t_r$ even for higher gas volume fractions. 
Furthermore, the estimated time of the bubble-thermistor contact remains much shorter than $t_r$ for $\alpha$ up to $5\%$ (see table \ref{table:t_bub}).
The present measurement technique is thus suitable for measurements of the temperature fluctuations in the liquid phase not only for $\alpha = 0.9 \%$, but also for higher gas volume fractions. 

 \begin{table}
 \begin{center}
\setlength\extrarowheight{6pt}

\begin{tabular}{   r  || r|  r  | r | r| r | r| r }
   {$\alpha$ $[\%]$}   &	{\ 0.5\ }	&{\ 	0.9\ }	&	{\ 2.0\ }	&	{\ 3.1\ }	&	{\ 3.9\ } &	{\ 5.0 \ }\\ 
   {$t_{2b}$ [ms]}  &	{\ 590\ }	&	{\ 400\ }	&	{\ 222\ }	&	{\ 167\ }	&	{\ 143\ }	&	{\ 123\ }\\ 
   $t_{int}$ [ms]	&	{\ 6.2\ }	&	{\ 7.2\ }	&	{\ 8.5\ }	&	{\ 9.8\ }	&	{\ 10.7\ }	&	{\ 11.7\ }\\  
   \end{tabular}
\end{center}
\caption{Measured mean time between bubble passages $t_{2b}$ and estimated time of bubble-thermistor contact $t_{int}$ for the studied gas volume fractions $\alpha$.}
\label{table:t_bub}
\end{table}

\section{Global heat transport enhancement} \label{global}

  \begin{figure}
\centering
\includegraphics[scale=1]{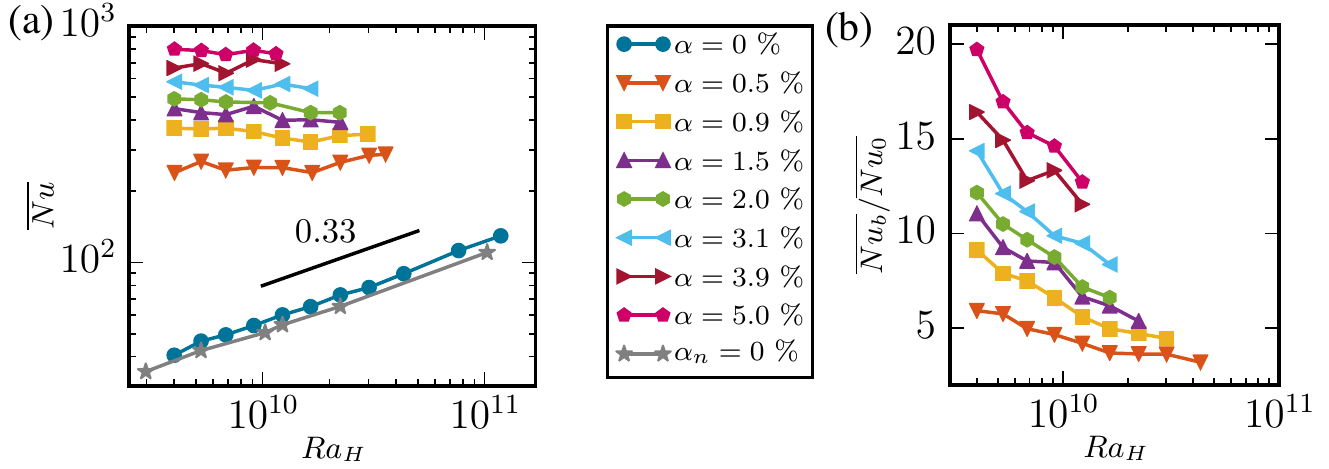}
\caption{(a) Dependence of Nusselt number $\overline{Nu}$ on the Rayleigh $Ra_H$ number for different gas volume fractions ($\alpha$ - experimental data, $\alpha_n$ - numerical simulations) - the size of the symbol corresponds to the error-bar, (b) Heat transfer enhancement, $\overline{Nu}_0$ - Nusselt number in single-phase case, $\overline{Nu}_b$ - Nusselt in bubbly flow.}
\label{Ra_Nu}
  \end{figure}

Let us now discuss the heat transport in the presence of a homogeneous bubble swarm for gas volume fraction $\alpha$ ranging from 0.5\% to 5\%, and the Rayleigh number ranging from $4.0 \times 10^9$ to utmost $3.6 \times 10^{10}$ (see Figure \ref{Ra_Nu}).
For the whole range of $\alpha$ and $Ra_H$, adding bubbles considerably increases the heat transport, since the Nusselt number is about an order of magnitude higher as compared to single-phase flow (see figure \ref{Ra_Nu} (a)). 
In order to better quantify the heat transport enhancement due to bubble injection, we show in figure \ref{Ra_Nu} (b) the ratio of the Nusselt number in the bubbly flow $\overline{Nu_b}$ to the Nusselt number in the single-phase $\overline{Nu_{0}}$ as a function of $Ra_H$ for different $\alpha$. 
We find that heat transfer is enhanced up to 20 times due to bubble injection, and that the enhancement increases with increasing $\alpha$ and decreasing $Ra_H$. 
Note that the decreasing trend of $\overline{Nu_b}/\overline{Nu_0}$ with $Ra_H$ occurs because the single-phase heat flux $\overline{Nu_0}$ increases with $Ra_H$.  

    \begin{figure}
\centering
\includegraphics[scale=1]{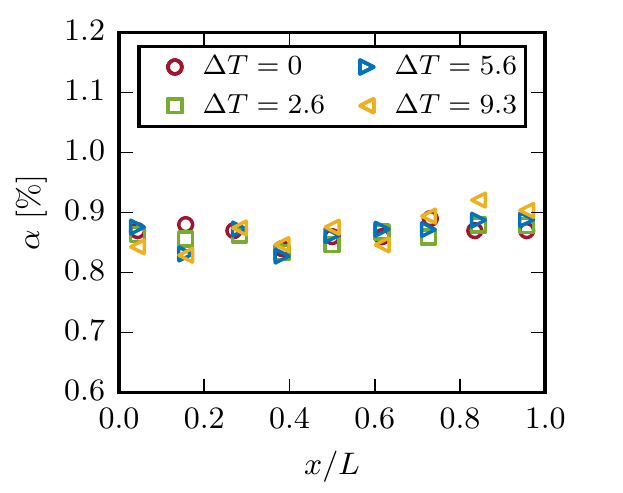}
\caption{The profile of the gas volume fraction for $\alpha = 0.9 \% $ at half height for different imposed temperature differences between hot and cold wall $\Delta T$.}
\label{gvf_heating}
    \end{figure}  

\begin{figure}
\centering
\includegraphics[scale=1]{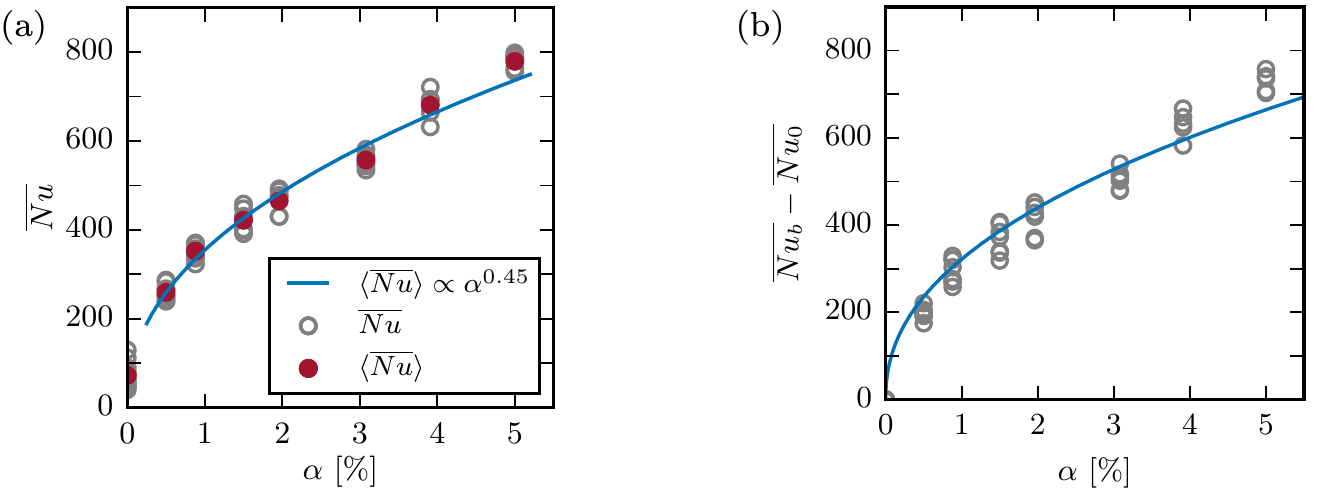}
\caption{(a) Dependence of Nusselt number $\overline{Nu}$ on gas volume fraction $\alpha$ (shallow gray circles present all the experimental measurements, red circles are values of Nusselt number averaged over the studied range of $Ra_H$ for each gas volume fraction); (b) The scaling of the difference between Nusselt number in bubbly flow $\overline{Nu}_b$ and Nusselt in  single-phase case $\overline{Nu}_0$ as a function of $\alpha$, blue curve corresponds to $\overline{Nu_b}-\overline{Nu_0} \propto \alpha^{0.45}$. }
\label{Nu_alpha}
    \end{figure}

Figure \ref{Ra_Nu} (a) also shows that for a fixed $\alpha \geq 0.5\%$  $\overline{Nu}$ remains nearly constant with increasing $Ra_H$. 
This indicates that the boundary layers developing along the walls are not limiting the heat transport anymore in the two-phase case.
Together with the observation that the heating does not induce a gradient in the gas volume fraction profile (see Figure \ref{gvf_heating}), this further implies that the temperature behaves as a passive scalar in bubbly flow. 
In order to understand the mechanism of the heat transport in bubbly flow we compare  the findings of our study to those of \cite{almeras2015mixing}, who showed that the transport of a passive scalar at high P\'eclet number ($Pe = V_{bub}d_{eq}/D \approx 10^6$, with $V_{bub}$ as the bubble velocity and $d_{eq}$ as the bubble diameter and $D$ as molecular diffusivity) by a homogeneous bubble swarm is a diffusive process. 
In the case of a diffusive process, the turbulent heat flux can be modeled as  $\overline{u'_{i} T'} = -D_{ii} \nabla \overline{T}$,  introducing the effective diffusivity  $D_{ii}$. 
If we take the heat transport to be a diffusive process in our study, the Nusselt number can be interpreted as the ratio between the effective diffusivity induced by the bubble swarm $D_{ii}$ and the thermal diffusivity $\chi$. 
Thus, from our measurements we can estimate the effective diffusivity induced by a bubble swarm for a gas volume fraction ranging from 0.5\% to 5$\%$. 
Note that since the temperature gradient is imposed in the horizontal direction, we assume that the measured effective diffusivity is  mainly in  the horizontal direction.
Figure \ref{Nu_alpha} (a) shows the Nusselt number $\langle \overline{Nu} \rangle $ averaged over the full range of Rayleigh number for a constant gas volume fraction as a function of $\alpha$. 
We clearly see that the averaged Nusselt number evolves as $\alpha^{0.45\pm0.025}$ .
Even if we subtract the single-phase Nusselt number (which might be thought of as the contribution of natural convection to the total heat transfer) from the one in bubbly flow the scaling remains unchanged (see figure \ref{Nu_alpha} (b)).
This trend is in a good agreement with the model of effective diffusivity proposed by \cite{almeras2015mixing}. 
In fact, the authors showed that at low gas volume fraction, the diffusion coefficient can be written as: $D_{ii} \propto u'\Lambda$, where  $u'$ is the standard deviation of the velocity fluctuations, and $\Lambda$ is the integral Lagrangian length scale ($\Lambda \simeq d_{bub}/C_{d0}$, $C_{d0} = 4d_{bub}g/3{V_0}^2$, here $C_{d0}$ is the drag coefficient and $V_0$ is the rise velocity of a single bubble \citep{riboux2010experimental}).
In the expression for the diffusion coefficient only the $u'$ depends on the gas volume fraction and this dependence is given by $u' \sim V_0 \alpha^{0.4} $ \citep{risso2002velocity}.  
In the present study, we expect to have a similar liquid agitation since bubble rising velocity and diameter are comparable. 
This yields the same evolution of the effective diffusivity $D_{ii}$ with the gas volume fraction $\alpha$ namely, $D_{ii} \propto \alpha^{0.45}$, extending the model proposed by \cite{almeras2015mixing} to lower P\'eclet number ($Pe\approx5000$ in this study). 
We also must stress here that the influence of the P\'eclet number on the effective diffusivity can be significant. 
In fact, numerical simulation performed by \cite{loisy2016direct}, at $\alpha = 2.4\%$ and $Re_b=30$ show that the effective diffusivity normalised by the molecular/thermal diffusivity varies linearly with the P\'eclet number (for $Pe$ ranging from $10^3$ to $10^6$). 
Consequently, since the P\'eclet number varies by three decades between the present study and the one of \cite{almeras2015mixing}, no quantitative comparison of the effective diffusivity can be performed. 
Therefore, further studies on the effect of the P\'eclet number on the effective diffusivity at high Reynolds number should be performed.

\section{Local characterisation of the heat transport} \label{local}
\begin{figure}
\centering
\includegraphics[scale=1]{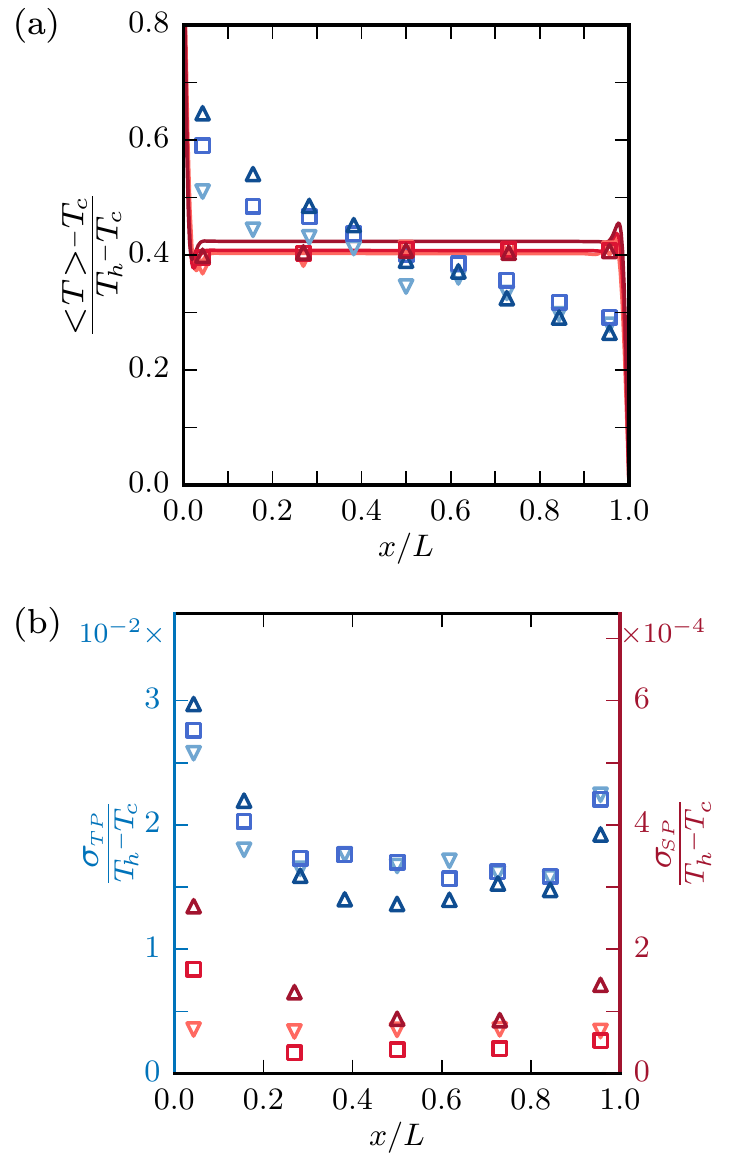}
\caption{(a) Normalised mean temperature profile at the mid-height (the overshoots close to the walls in the numerical data appear because the thin layer of warmed up (cooled down) fluids moving upward (downward) are still colder (hotter) than the bulk where there is a stable stratification), (b) Normalised standard deviation of the temperature fluctuations. Red lines (numerical results in (a)) and symbols (experimental data) present single-phase, blue symbols present two-phase with $\alpha = 0.9\%$, for various Rayleigh numbers: $Ra_H = 5.2 \times 10^{9}$ (downwards triangles), $Ra_H = 1.6 \times 10^{10}$ (squares), and $Ra_H = 2.2 \times 10^{10}$ (upward triangles)).}
\label{t_profile}
	\end{figure}

\begin{figure}
\centering
\includegraphics[scale=1]{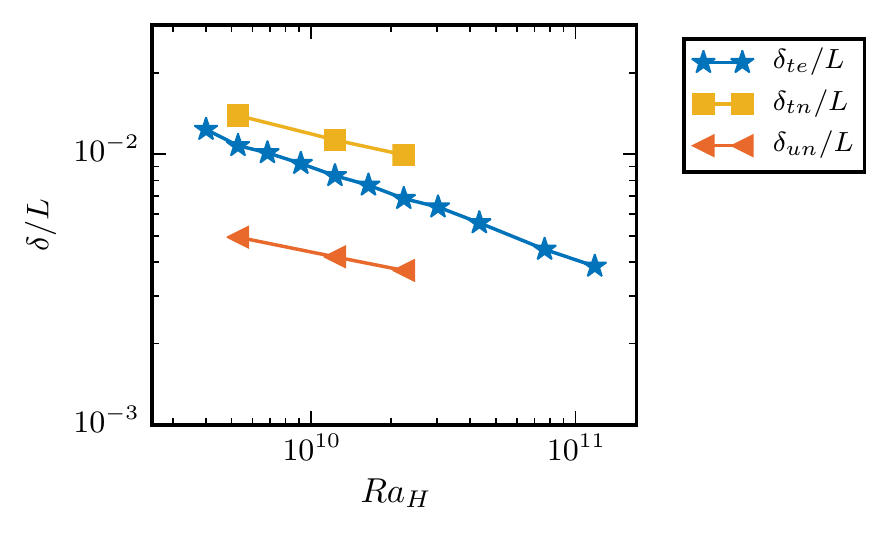}
\caption{Normalised boundary layer thickness in the single phase as a function of $Ra_H$. Here ${\delta _u}_n$ and ${\delta _t}_n$ are numerically obtained thickness of the kinetic and thermal boundary layer, respectively. Thickness of the thermal boundary layer obtained from the experiments is given as ${\delta _t}_e$. }
\label{BL2}
    \end{figure}

In order to gain further insight into the heat transport enhancement, we performed local liquid temperature measurements by traversing the thermistor along the length of the setup at mid-height for three Rayleigh numbers ($Ra_H = 5.2 \times 10^{9}$, $Ra_H = 1.6 \times 10^{10}$, and $Ra_H = 2.2 \times 10^{10}$), and for $\alpha = 0\%$ and $\alpha = 0.9\%$.
Figure \ref{t_profile}~(a) shows the normalised temperature profiles. 
For the single-phase cases, as expected from the range of $Ra_H$ and the $H/L$ in our study, a flat temperature profile along the length is observed in the bulk at mid-height \citep{elder1965turbulent,markatos1984laminar,bejan1984boundary,belmonte1994temperature,ng2015vertical,shishkina2016thermal,ng2017changes}.
After normalisation, the single-phase temperature profiles for all three Rayleigh numbers overlap.
However, due to present heat losses to the outside at the top of the setup since the setup is open on the top, the temperature profiles do not collapse at $\frac{\langle T \rangle - T_c}{T_h-Tc} =0.5$ but at $\frac{\langle T \rangle - T_c}{T_h-Tc} =0.4$.
This has to be taken into account when comparing numerical data with the data obtained experimentally. 
Therefore, after shifting the numerically obtained temperature profiles good agreement is found between the two.
The spatial temperature gradient in the single-phase case is located in the thermal boundary layer whose thickness $\delta_t$ is estimated from the numerical data to be $\mathcal{O}(\unit{1}{\milli \meter})$. 
In figure \ref{BL2} we show the boundary layer thickness in the single phase case as a function of $Ra_H$.
Note that our flow configuration is different from the one present in classical Rayleigh-B{\'e}nard setup. 
Here the thermal boundary layer is defined as wall distance to the intercept of $\overline{T}~=~T_h~+~d\overline{T}/dx |_w~x$ and $\overline{T} = T_h - \Delta T/2$ and the kinetic boundary layer is given as an intercept of $\overline{u}=d\overline{u}/dx|_w x$ and $\overline{u} = \overline{u}_{max}$ (see e.g. \cite{ng2015vertical} for more details).
The thickness of the thermal boundary layer based on the experimentally obtained Nusselt number is comparable to the one obtained numerically (see Figure \ref{BL2}).

As seen in figure \ref{t_profile} (a), the mean temperature profiles in the case of bubbly flow is completely different from that of single phase flow. 
In the bulk, two known mixing mechanisms contribute to the distortion of the flat temperature profile that was observed in the single-phase case: (i) capture and transport by the bubble wakes \citep{bouche2013mixing}, and (ii) dispersion by the bubble-induced turbulence which is the dominant one as shown by \citet{almeras2015mixing}. 
Near the heated and cooled walls, we visually observed the bubbles bouncing along the walls. 
This presumably disturbs the thermal boundary layers; however, this cannot be measured due to insufficient experimental resolution.

Figures \ref{t_profile} (b) and \ref{fig:fiber_temp} (a) and \ref{fig:fiber_temp} (b) show that the temperature fluctuations induced by bubbles are even two orders of magnitudes higher than in the single-phase case ($T'= T - \langle T \rangle$, where $T'$ is the temperature fluctuations, $T$ is the measured instantaneous temperature and $ \langle T \rangle$ is the time averaged temperature at the measurement point).
The normalised standard deviations of the fluctuations in both single-phase and two-phase are higher closer to the cold and hot walls than in the center of the setup.
This is possibly due to the temperature probe seeing more hot and cold plumes closer to the heated and cold walls, a well known phenomenon from Rayleigh-B\'{e}nard flow \citep{ahlers2009heat}. 
Here slight asymmetry of the temperature profiles and the profiles of normalised standard deviation close to the walls must be attributed to the difference in the nature of heating and cooling of the sidewalls~(see asymmetry also in figure \ref{t_profile} (a)). 
Figures \ref{fig:fiber_temp} (a) and (b) also demonstrate that the time scales of fluctuations in single-phase and two-phase are different, along with much more intense temperature fluctuations for the bubbly flow.

\begin{figure}
\centering
\includegraphics[scale=1]{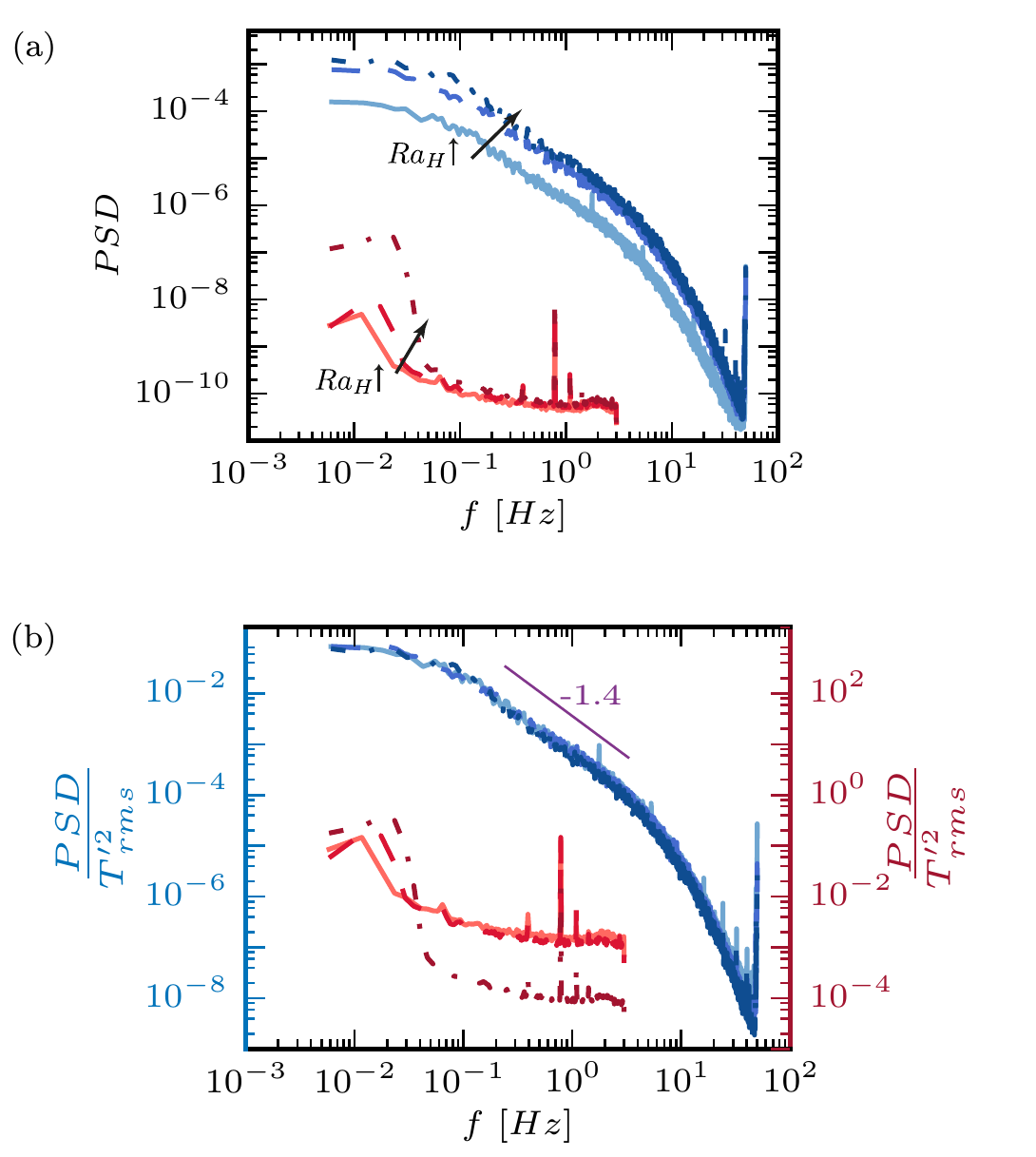}
\caption{Raw (a) and  normalised (b) power spectra of the temperature fluctuations at the center of the setup for single-phase (red lines) and two-phase $\alpha = 0.9\%$ (blue lines); solid line: $Ra_H = 5.2 \times 10^{9}$; dashed line: $Ra_H = 1.6 \times 10^{10}$;  dash-dotted line: $Ra_H = 2.2 \times 10^{10}$.}  
\label{PSD_center}
\end{figure}

To explore this in better detail, we now present the power spectrum of the temperature fluctuations (``thermal power spectrum'') for both cases.
Figure \ref{PSD_center} (a) shows this power spectrum of temperature fluctuations at mid-height in the centre of the setup for the single- and two-phase cases.
In the single-phase case the measured temperature fluctuations are limited to frequencies lower than $10^{-1}$ Hz.
At around $10^{-2}$ Hz we observe a peak, beyond which there is a very steep decrease of the spectrum (the same frequency can be observed in the figure \ref{fig:fiber_temp} a)).
As well known \citep{castaing1989scaling} this peak corresponds to the large scale circulation frequency ($f_{LS} \approx V_{ff}/4H$), which can be estimated from the free fall velocity  $V_{ff} = \sqrt{g \beta \Delta T H }$ which is $\sim \unit{6}{\centi \meter \per \second}$ for lowest $Ra_H$ and $ \sim \unit{11}{\centi \meter \per \second}$ for the highest $Ra_H$.
In both single-phase and two-phase cases, a higher level of thermal power is seen for higher $Ra_H$ numbers. 
The same trend is seen for all the measurement positions at mid-height. 
If we now compare the single-phase and two-phase spectra, we can see that with the bubble injection, the thermal power of the fluctuations is increased by nearly three orders of magnitude. 
The bubbly flow also shows fluctuations at a range of time scales, with a gradual decay of thermal power from $f \simeq \unit{0.1}{\hertz} - \unit{3}{\hertz}$.
The observation that substantial power of the temperature fluctuations resides at smaller time scales, as compared to the single-phase where the power mainly resides at the largest time scales,  further confirms that the bubble-induced liquid fluctuations are the dominant contribution to the total heat transfer.

The thermal power spectra plots in figure \ref{PSD_center} (a) show a $Ra_H$ dependence for both single- and two-phase cases. 
While upon normalising with the scale of temperature fluctuations $T_{rms}^2$ in the single phase we do not observe complete overlap of the spectra possibly due to noise present at higher frequencies, in bubbly flow we observe a nearly perfect collapse of the three Rayleigh numbers ~(see figure~\ref{PSD_center}(b)). 
This suggests a universal behavior for bubbly flow.  
Interestingly, this is similar to the velocity fluctuations spectra observed for bubbly flows~\citep{lance1991turbulence, riboux2010experimental,roghair2011energy, prakash2016energy}, where a normalisation with the scale of velocity fluctuations ${u^2_{rms}}$ demonstrates universality. 
Furthermore, the same behavior is seen at all measurement positions and all Rayleigh numbers (note that in figure \ref{PSD_center} (b), we have shown the measurements at the centre only). 
We also observe a clear slope of $-1.4$ at the scales  $f \simeq \unit{0.1}{\hertz}  - \unit{3}{\hertz}$.
It remains unclear why this exact slope is present, and how it can be attributed to bubble-induced turbulence.
Events occurring at shorter time scales, such as at frequencies where the $-3$ slope is present in velocity spectra for bubble-induced turbulence, typically starting at $1/t_{pseudo} \sim 35 $ Hz~\citep{riboux2010experimental}, would be undetectable here due to the limiting response time of the thermistor used.

\section{Summary of main results and discussion} \label{summary}
\begin{figure}
\centering
\includegraphics[scale=1]{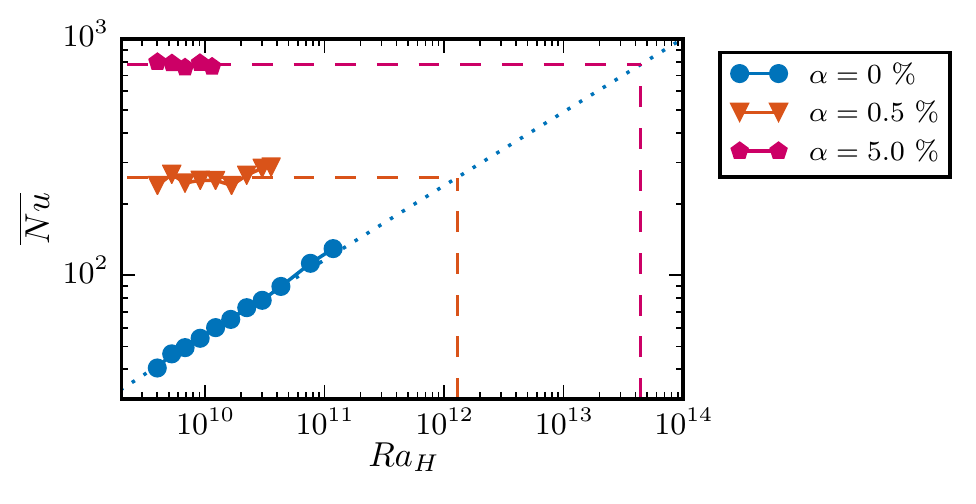}
\caption{Extrapolation of the Nusselt number to high Rayleigh numbers for the two-phase (dashed lines - lowest and highest studied $\alpha$), and the single phase case (dotted line). A crossover between the extrapolated single- and two-phase cases occurs at $Ra_{H} \approx 1.2 \times 10^{12}$ for $\alpha = 0.5\% $, and at $Ra_{H} \approx 4\times 10^{13}$ for $\alpha = 5\%$.  However, as we approach these $Ra_H$, we expect the Rayleigh-independent trends of Nusselt number to change, namely to increase with increasing $Ra_H$.}
\label{fig:Ra_Nu_ext}
\end{figure}

An experimental study on heat transport in homogeneous bubbly flow has been conducted. 
The experiments are performed in a rectangular bubble column heated from one side and cooled from the other (see figure \ref{fig1}).
Two parameters are varied: the gas volume fraction and the Rayleigh number.
The gas volume fraction ranges from $0\%$ to $5\%$, and the bubble diameters are around $\unit{2.5}{\milli \meter}$. 
The Rayleigh number is in the range $4.0 \times 10^9 - 1.2 \times 10^{11}$.

First, we focus on characterization of the global heat transfer for single-phase and two-phase cases. 
We find that two completely different mechanisms govern the heat transport in these two cases.
In the single-phase case, the vertical natural convection is driven solely by the imposed difference between the mean wall temperatures. 
In this configuration the temperature acts as an active scalar driving the flow.
The Nusselt number increases with increasing Rayleigh number, and as expected effectively scales as: $\overline{Nu} \sim Ra_H^{0.33}$ (see figure \ref{Ra_Nu} (a)). 
However, in the case of homogeneous bubbly flow the heat transfer comes from two different contributions: natural convection driven by the horizontal temperature gradient and the bubble induced diffusion, where the latter dominates.
This is substantiated by our observations that the Nusselt number in bubbly flow is nearly independent of the Rayleigh number and depends solely on the gas volume fraction, evolving as: $\overline{Nu}\propto \alpha^{0.45}$ (see figure \ref{Nu_alpha}).
We thus find nearly the same scaling as in the case of the mixing of a passive tracer in a homogeneous bubbly flow for a low gas volume fraction (\cite{almeras2015mixing}), which implies that the bubble-induced mixing is indeed limiting the efficiency of the heat transfer.

We further performed local temperature measurements at the mid-height of the setup for the gas volume fraction of $\alpha = 0\%$ and $\alpha = 0.9\%$, and for Rayleigh numbers  $5.2 \times 10^{9}$, $1.6 \times 10^{10}$ and $2.2 \times 10^{10}$.
For single-phase flow, we observe that the mean temperature remains constant in the bulk at mid-height. 
However, in the two-phase flow case, this is completely obstructed by the mixing induced by bubbles.
Injection of bubbles induces up to 200 times stronger temperature fluctuations (see figure \ref{t_profile} (b)).
These fluctuations over a wide spectrum of frequencies (see figure \ref{PSD_center}) are thus the signature of the heat transport enhancement due to bubble injection.
A clear slope of $-1.4$ at the scales $f \simeq \unit{0.1}{\hertz} - \unit{3}{\hertz}$ was also observed. 
In order to understand why this slope is present in that range it would be beneficial to perform local velocity measurements in the flow with heating, which is objective of our future studies.
Further examination with fully resolved numerical simulations will also help us understand the effect of bubbles.
These simulations are planned for future work, as well.

To conclude, we observe up to 20 times heat transfer enhancement due to bubble injection (see figure \ref{Ra_Nu} (b)). 
This demonstrates that the diffusion induced by bubbles is a highly effective mechanism for heat transfer enhancement.
Nevertheless, several questions remain unanswered.
One question of great practical importance is: at what Rayleigh number will the contribution of natural convection to the total heat transfer become comparable or even greater than the contribution of bubble-induced turbulence?
If we extrapolate the data to higher Rayleigh numbers, we can obtain the maximum expected value of $Ra_H$ at which the two contributions are comparable (see dashed lines in figure \ref{fig:Ra_Nu_ext}).
This occurs at $Ra_{H} \approx 1.2 \times 10^{12}$ for $\alpha = 0.5\% $ and at $Ra_{H} \approx 4\times 10^{13}$ for $\alpha = 5\% $.
As we approach these $Ra_H$, we expect the Rayleigh-independent behavior of Nusselt number to change. 
Presumably, when $Ra_H$ is sufficiently large, the Nusselt number will increase with $Ra_H$ even for the two-phase cases. 
Based on our current knowledge, it is difficult to predict at what $Ra_H$ this trend will change. This calls for future investigations spanning a wider range of control parameters.


We thank G.-W. Bruggert, M. Bos, and B. Benschop for technical support and Y.-C. Xie for the help with the experimental technique.
This work is part of the Industrial Partnership Programme i36 Dense Bubbly Flows that is carried out under an agreement between Akzo Nobel Chemicals International B.V., DSM Innovation Center B.V., SABIC Global Technologies B.V., Shell Global Solutions B.V., Tata Steel Nederland Technology B.V. and the Netherlands Organisation for Scientific Research (NWO). We also thank STW foundation of the Netherlands, European High-Performance Infrastructures in Turbulence (EuHIT), and COST Action MP1305 for support. This work was also supported by The Netherlands Center for Multiscale Catalytic Energy Conversion (MCEC), an NWO Gravitation Programme funded by the Ministry of Education, Culture and Science of the government of The Netherlands. Chao Sun acknowledges the financial support from Natural Science Foundation of China under Grant No. 11672156.


\begin{thebibliography}{50}
\expandafter\ifx\csname natexlab\endcsname\relax\def\natexlab#1{#1}\fi
\def\au#1{#1} \def\ed#1{#1} \def\yr#1{#1}\def\at#1{#1}\def\jt#1{\textit{#1}}
  \def\bt#1{#1}\def\bvol#1{\textbf{#1}} \def\vol#1{#1} \def\pg#1{#1}
  \def\publ#1{#1}\def\arxiv#1{#1}\def\org#1{#1}\def\st#1{\textit{#1}}

\bibitem[Ahlers {\em et~al.\/}(2009)Ahlers, Grossmann \& Lohse]{ahlers2009heat}
{\sc \au{Ahlers, G.}, \au{Grossmann, S.} \& \au{Lohse, D.}} \yr{2009}  \at{Heat
  transfer and large scale dynamics in turbulent {R}ayleigh-{B}{\'e}nard
  convection}.  \jt{Rev. Mod. Phys.}  \bvol{81}~(2),  \pg{503}.

\bibitem[Alm{\'e}ras(2014)]{almeras2014etude}
{\sc \au{Alm{\'e}ras, E.}} \yr{2014}  \at{{\'E}tude des propri{\'e}t{\'e}s de
  transport et de m{\'e}lange dans les {\'e}coulements {\`a} bulles}. PhD
  thesis, Universit\'{e} de Toulouse.

\bibitem[Alm{\'e}ras {\em et~al.\/}(2017)Alm{\'e}ras, Mathai, Lohse \&
  Sun]{almeras2017experimental}
{\sc \au{Alm{\'e}ras, E.}, \au{Mathai, V.}, \au{Lohse, D.} \& \au{Sun, C.}}
  \yr{2017}  \at{Experimental investigation of the turbulence induced by a
  bubble swarm rising within an incident turbulence}.  \jt{arXiv preprint
  arXiv:1706.02593} .

\bibitem[Alm{\'e}ras {\em et~al.\/}(2015)Alm{\'e}ras, Risso, Roig, Cazin, Plais
  \& Augier]{almeras2015mixing}
{\sc \au{Alm{\'e}ras, E.}, \au{Risso, F.}, \au{Roig, V.}, \au{Cazin, S.},
  \au{Plais, C.} \& \au{Augier, F.}} \yr{2015}  \at{Mixing by bubble-induced
  turbulence}.  \jt{J. Fluid Mech.}  \bvol{776},  \pg{458--474}.

\bibitem[Bejan(1984)]{bejan1984boundary}
{\sc \au{Bejan, A.}} \yr{1984}  \at{The boundary layer natural convection
  regime in a rectangular cavity with uniform heat flux from the side}.  \jt{J.
  Heat Transfer}  \bvol{106},  \pg{99}.

\bibitem[Bejan(2004)]{bejan2013convection}
{\sc \au{Bejan, A.}} \yr{2004} {\em Convection heat transfer\/}.  \publ{John
  Wiley \& Sons}.

\bibitem[Belmonte {\em et~al.\/}(1994)Belmonte, Tilgner \&
  Libchaber]{belmonte1994temperature}
{\sc \au{Belmonte, A.}, \au{Tilgner, A.} \& \au{Libchaber, A.}} \yr{1994}
  \at{Temperature and velocity boundary layers in turbulent convection}.
  \jt{Phys. Rev. E}  \bvol{50}~(1),  \pg{269}.

\bibitem[Bouche {\em et~al.\/}(2013)Bouche, Cazin, Roig \&
  Risso]{bouche2013mixing}
{\sc \au{Bouche, E.}, \au{Cazin, S.}, \au{Roig, V.} \& \au{Risso, F.}}
  \yr{2013}  \at{Mixing in a swarm of bubbles rising in a confined cell
  measured by mean of {P}{L}{I}{F} with two different dyes}.  \jt{Exp. Fluids}
  \bvol{54}~(6),  \pg{1552}.

\bibitem[Castaing {\em et~al.\/}(1989)Castaing, Gunaratne, Heslot, Kadanoff,
  Libchaber, Thomae, Wu, Zaleski \& Zanetti]{castaing1989scaling}
{\sc \au{Castaing, B.}, \au{Gunaratne, G.}, \au{Heslot, F.}, \au{Kadanoff, L.},
  \au{Libchaber, A.}, \au{Thomae, S.}, \au{Wu, X.-Z.}, \au{Zaleski, S.} \&
  \au{Zanetti, G.}} \yr{1989}  \at{Scaling of hard thermal turbulence in
  {R}ayleigh-{B}{\'e}nard convection}.  \jt{J. Fluid Mech.}  \bvol{204},
  \pg{1--30}.

\bibitem[Chong {\em et~al.\/}(2015)Chong, Huang, Kaczorowski \&
  Xia]{chong2015condensation}
{\sc \au{Chong, K.~L.}, \au{Huang, S.}, \au{Kaczorowski, M.} \& \au{Xia,
  K.-Q.}} \yr{2015}  \at{Condensation of coherent structures in turbulent
  flows}.  \jt{Phys. Rev. Lett.}  \bvol{115}~(26),  \pg{264503}.

\bibitem[Dabiri \& Tryggvason(2015)]{dabiri2015heat}
{\sc \au{Dabiri, S.} \& \au{Tryggvason, G.}} \yr{2015}  \at{Heat transfer in
  turbulent bubbly flow in vertical channels}.  \jt{Chem. Eng. Sci.}
  \bvol{122},  \pg{106--113}.

\bibitem[Deckwer(1980)]{deckwer1980mechanism}
{\sc \au{Deckwer, W.-D.}} \yr{1980}  \at{On the mechanism of heat transfer in
  bubble column reactors}.  \jt{Chem. Eng. Sci.}  \bvol{35}~(6),
  \pg{1341--1346}.

\bibitem[Deen \& Kuipers(2013)]{deen2013direct}
{\sc \au{Deen, N.~G.} \& \au{Kuipers, J. A.~M.}} \yr{2013}  \at{Direct
  numerical simulation of wall-to liquid heat transfer in dispersed gas-liquid
  two-phase flow using a volume of fluid approach}.  \jt{Chem. Eng. Sci.}
  \bvol{102},  \pg{268--282}.

\bibitem[Elder(1965)]{elder1965turbulent}
{\sc \au{Elder, J.~W.}} \yr{1965}  \at{Turbulent free convection in a vertical
  slot}.  \jt{J. Fluid Mech.}  \bvol{23}~(1),  \pg{99--111}.

\bibitem[van Gils {\em et~al.\/}(2013)van Gils, Narezo~Guzman, Sun \&
  Lohse]{van2013importance}
{\sc \au{van Gils, D. P.~M.}, \au{Narezo~Guzman, D.}, \au{Sun, C.} \&
  \au{Lohse, D.}} \yr{2013}  \at{The importance of bubble deformability for
  strong drag reduction in bubbly turbulent {T}aylor--{C}ouette flow}.  \jt{J.
  Fluid Mech.}  \bvol{722},  \pg{317--347}.

\bibitem[Kitagawa {\em et~al.\/}(2008)Kitagawa, Kosuge, Uchida \&
  Hagiwara]{kitagawa2008heat}
{\sc \au{Kitagawa, A.}, \au{Kosuge, K.}, \au{Uchida, K.} \& \au{Hagiwara, Y.}}
  \yr{2008}  \at{Heat transfer enhancement for laminar natural convection along
  a vertical plate due to sub-millimeter-bubble injection}.  \jt{Exp. Fluids}
  \bvol{45}~(3),  \pg{473--484}.

\bibitem[Kitagawa \& Murai(2013)]{kitagawa2013natural}
{\sc \au{Kitagawa, A.} \& \au{Murai, Y.}} \yr{2013}  \at{Natural convection
  heat transfer from a vertical heated plate in water with microbubble
  injection}.  \jt{Chem. Eng. Sci.}  \bvol{99},  \pg{215--224}.

\bibitem[Kitagawa {\em et~al.\/}(2009)Kitagawa, Uchida \&
  Hagiwara]{kitagawa2009effects}
{\sc \au{Kitagawa, A.}, \au{Uchida, K.} \& \au{Hagiwara, Y.}} \yr{2009}
  \at{Effects of bubble size on heat transfer enhancement by sub-millimeter
  bubbles for laminar natural convection along a vertical plate}.  \jt{Int. J.
  Heat Fluid Flow}  \bvol{30}~(4),  \pg{778--788}.

\bibitem[Lakkaraju {\em et~al.\/}(2013)Lakkaraju, Stevens, Oresta, Verzicco,
  Lohse \& Prosperetti]{lakkaraju2013heat}
{\sc \au{Lakkaraju, R.}, \au{Stevens, R.}, \au{Oresta, P.}, \au{Verzicco, R.},
  \au{Lohse, D.} \& \au{Prosperetti, A.}} \yr{2013}  \at{Heat transport in
  bubbling turbulent convection}.  \jt{Proc. Nat. Acad. Sci.}  \bvol{110}~(23),
   \pg{9237--9242}.

\bibitem[Lance \& Bataille(1991)]{lance1991turbulence}
{\sc \au{Lance, M.} \& \au{Bataille, J.}} \yr{1991}  \at{Turbulence in the
  liquid phase of a uniform bubbly air--water flow}.  \jt{J. Fluid Mech.}
  \bvol{222},  \pg{95--118}.

\bibitem[Lohse \& Xia(2010)]{lohse2010small}
{\sc \au{Lohse, D.} \& \au{Xia, K.-Q.}} \yr{2010}  \at{Small-scale properties
  of turbulent {R}ayleigh-{B}{\'e}nard convection}.  \jt{Annu. Rev. Fluid
  Mech.}  \bvol{42},  \pg{335--364}.

\bibitem[Loisy(2016)]{loisy2016direct}
{\sc \au{Loisy, A.}} \yr{2016}  \at{Direct numerical simulation of bubbly
  flows: coupling with scalar transport and turbulence}. PhD thesis,
  Universit{\'e} de Lyon.

\bibitem[Markatos \& Pericleous(1984)]{markatos1984laminar}
{\sc \au{Markatos, N.~C.} \& \au{Pericleous, K.~A.}} \yr{1984}  \at{Laminar and
  turbulent natural convection in an enclosed cavity}.  \jt{Int. J. Heat Mass
  Transfer}  \bvol{27}~(5),  \pg{755--772}.

\bibitem[Mercado~Mart{\'\i}nez {\em et~al.\/}(2010)Mercado~Mart{\'\i}nez,
  Chehata~G{\'\o}mez, Van~Gils, Sun \& Lohse]{mercado2010bubble}
{\sc \au{Mercado~Mart{\'\i}nez, J.}, \au{Chehata~G{\'\o}mez, D.}, \au{Van~Gils,
  D.}, \au{Sun, C.} \& \au{Lohse, D.}} \yr{2010}  \at{On bubble clustering and
  energy spectra in pseudo-turbulence}.  \jt{J. Fluid Mech.}  \bvol{650},
  \pg{287--306}.

\bibitem[Narezo~Guzman {\em et~al.\/}(2016{\natexlab{{\em a\/}}})Narezo~Guzman,
  Fraczek, Reetz, Sun, Lohse \& Ahlers]{guzman2016vapour}
{\sc \au{Narezo~Guzman, D.}, \au{Fraczek, T.}, \au{Reetz, C.}, \au{Sun, C.},
  \au{Lohse, D.} \& \au{Ahlers, G.}} \yr{2016{\natexlab{{\em a\/}}}}
  \at{Vapour--bubble nucleation and dynamics in turbulent
  {R}ayleigh--{B}{\'e}nard convection}.  \jt{J. Fluid Mech.}  \bvol{795},
  \pg{60--95}.

\bibitem[Narezo~Guzman {\em et~al.\/}(2016{\natexlab{{\em b\/}}})Narezo~Guzman,
  Xie, Chen, Rivas, Sun, Lohse \& Ahlers]{guzman2016heat}
{\sc \au{Narezo~Guzman, D.}, \au{Xie, Y.}, \au{Chen, S.}, \au{Rivas, D.~F.},
  \au{Sun, C.}, \au{Lohse, D.} \& \au{Ahlers, G.}} \yr{2016{\natexlab{{\em
  b\/}}}}  \at{Heat--flux enhancement by vapour--bubble nucleation in
  {R}ayleigh--{B}{\'e}nard turbulence}.  \jt{J. Fluid Mech.}  \bvol{787},
  \pg{331--366}.

\bibitem[Ng {\em et~al.\/}(2015)Ng, Ooi, Lohse \& Chung]{ng2015vertical}
{\sc \au{Ng, C.~S.}, \au{Ooi, A.}, \au{Lohse, D.} \& \au{Chung, D.}} \yr{2015}
  \at{Vertical natural convection: application of the unifying theory of
  thermal convection}.  \jt{J. Fluid Mech.}  \bvol{764},  \pg{349--361}.

\bibitem[Ng {\em et~al.\/}(2017)Ng, Ooi, Lohse \& Chung]{ng2017changes}
{\sc \au{Ng, C.~S.}, \au{Ooi, A.}, \au{Lohse, D.} \& \au{Chung, D.}} \yr{2017}
  \at{Changes in the boundary-layer structure at the edge of the ultimate
  regime in vertical natural convection}.  \jt{J. Fluid Mech.}  \bvol{825},
  \pg{550--572}.

\bibitem[Oresta {\em et~al.\/}(2009)Oresta, Verzicco, Lohse \&
  Prosperetti]{oresta2009heat}
{\sc \au{Oresta, P.}, \au{Verzicco, R.}, \au{Lohse, D.} \& \au{Prosperetti,
  A.}} \yr{2009}  \at{Heat transfer mechanisms in bubbly
  {R}ayleigh-{B}{\'e}nard convection}.  \jt{Phys. Rev. E}  \bvol{80}~(2),
  \pg{026304}.

\bibitem[van~der Poel {\em et~al.\/}(2015)van~der Poel, Ostilla-M{\'o}nico,
  Donners \& Verzicco]{van2015pencil}
{\sc \au{van~der Poel, E.~P.}, \au{Ostilla-M{\'o}nico, R.}, \au{Donners, J.} \&
  \au{Verzicco, R.}} \yr{2015}  \at{A pencil distributed finite difference code
  for strongly turbulent wall--bounded flows}.  \jt{Computers \& Fluids}
  \bvol{116},  \pg{10--16}.

\bibitem[Prakash {\em et~al.\/}(2016)Prakash, Mart{\'\i}nez~Mercado, van
  Wijngaarden, Mancilla, Tagawa, Lohse \& Sun]{prakash2016energy}
{\sc \au{Prakash, V.~N.}, \au{Mart{\'\i}nez~Mercado, J.}, \au{van Wijngaarden,
  L.}, \au{Mancilla, E.}, \au{Tagawa, Y.}, \au{Lohse, D.} \& \au{Sun, C.}}
  \yr{2016}  \at{Energy spectra in turbulent bubbly flows}.  \jt{J. Fluid
  Mech.}  \bvol{791},  \pg{174--190}.

\bibitem[Rensen {\em et~al.\/}(2005)Rensen, Luther \& Lohse]{rensen2005effect}
{\sc \au{Rensen, J.}, \au{Luther, S.} \& \au{Lohse, D.}} \yr{2005}  \at{The
  effect of bubbles on developed turbulence}.  \jt{J. Fluid Mech.}  \bvol{538},
   \pg{153--187}.

\bibitem[Riboux {\em et~al.\/}(2010)Riboux, Risso \&
  Legendre]{riboux2010experimental}
{\sc \au{Riboux, G.}, \au{Risso, F.} \& \au{Legendre, D.}} \yr{2010}
  \at{Experimental characterization of the agitation generated by bubbles
  rising at high {R}eynolds number}.  \jt{J. Fluid Mech.}  \bvol{643},
  \pg{509--539}.

\bibitem[Risso \& Ellingsen(2002)]{risso2002velocity}
{\sc \au{Risso, F.} \& \au{Ellingsen, K.}} \yr{2002}  \at{Velocity fluctuations
  in a homogeneous dilute dispersion of high-{R}eynolds-number rising bubbles}.
   \jt{J. Fluid Mech.}  \bvol{453},  \pg{395--410}.

\bibitem[Roche {\em et~al.\/}(2001)Roche, Castaing, Chabaud \&
  H{\'e}bral]{roche2001observation}
{\sc \au{Roche, P.-E.}, \au{Castaing, B.}, \au{Chabaud, B.} \& \au{H{\'e}bral,
  B.}} \yr{2001}  \at{Observation of the 1/2 power law in
  {R}ayleigh-{B}{\'e}nard convection}.  \jt{Phys. Rev. E}  \bvol{63}~(4),
  \pg{045303}.

\bibitem[Roghair {\em et~al.\/}(2011)Roghair, Martinez~Mercado, Van
  Sint~Annaland, Kuipers, Sun \& Lohse]{roghair2011energy}
{\sc \au{Roghair, I.}, \au{Martinez~Mercado, J.}, \au{Van Sint~Annaland, M.},
  \au{Kuipers, H.}, \au{Sun, C.} \& \au{Lohse, D.}} \yr{2011}  \at{Energy
  spectra and bubble velocity distributions in pseudo-turbulence: {N}umerical
  simulations vs. experiments}.  \jt{Int. J. Multiphase Flow}  \bvol{37}~(9),
  \pg{1093--1098}.

\bibitem[Sato {\em et~al.\/}(1981{\natexlab{{\em a\/}}})Sato, Sadatomi \&
  Sekoguchi]{sato1981momentum}
{\sc \au{Sato, Y.}, \au{Sadatomi, M.} \& \au{Sekoguchi, K.}}
  \yr{1981{\natexlab{{\em a\/}}}}  \at{Momentum and heat transfer in two-phase
  bubble flow—i. {T}heory}.  \jt{Int. J. Multiphase Flow}  \bvol{7}~(2),
  \pg{167--177}.

\bibitem[Sato {\em et~al.\/}(1981{\natexlab{{\em b\/}}})Sato, Sadatomi \&
  Sekoguchi]{sato1981momentum2}
{\sc \au{Sato, Y.}, \au{Sadatomi, M.} \& \au{Sekoguchi, K.}}
  \yr{1981{\natexlab{{\em b\/}}}}  \at{Momentum and heat transfer in two-phase
  bubble flow—ii. {A} comparison between experimental data and theoretical
  calculations}.  \jt{Int. J. Multiphase Flow}  \bvol{7}~(2),  \pg{179--190}.

\bibitem[Schmidt {\em et~al.\/}(2011)Schmidt, Oresta, Toschi, Verzicco, Lohse
  \& Prosperetti]{schmidt2011modification}
{\sc \au{Schmidt, L.~E.}, \au{Oresta, P.}, \au{Toschi, F.}, \au{Verzicco, R.},
  \au{Lohse, D.} \& \au{Prosperetti, A.}} \yr{2011}  \at{Modification of
  turbulence in {R}ayleigh--{B}{\'e}nard convection by phase change}.  \jt{New
  J. Phys.}  \bvol{13}~(2),  \pg{025002}.

\bibitem[Sekoguchi {\em et~al.\/}(1980)Sekoguchi, Nakazatomi \&
  Tanaka]{sekoguch1980forced}
{\sc \au{Sekoguchi, K.}, \au{Nakazatomi, M.} \& \au{Tanaka, O.}} \yr{1980}
  \at{Forced convective heat transfer in vertical air-water bubble flow}.
  \jt{Bulletin of JSME}  \bvol{23}~(184),  \pg{1625--1631}.

\bibitem[Shakerin {\em et~al.\/}(1988)Shakerin, Bohn \&
  Loehrke]{shakerin1988natural}
{\sc \au{Shakerin, S.}, \au{Bohn, M.} \& \au{Loehrke, R.}} \yr{1988}
  \at{Natural convection in an enclosure with discrete roughness elements on a
  vertical heated wall}.  \jt{Int. J. Multiphase Flow}  \bvol{31}~(7),
  \pg{1423--1430}.

\bibitem[Shishkina \& Horn(2016)]{shishkina2016thermal}
{\sc \au{Shishkina, O.} \& \au{Horn, S.}} \yr{2016}  \at{Thermal convection in
  inclined cylindrical containers}.  \jt{J. Fluid Mech.}  \bvol{790}.

\bibitem[Spandan {\em et~al.\/}(2016)Spandan, Ostilla-M{\'o}nico, Verzicco \&
  Lohse]{spandan2016drag}
{\sc \au{Spandan, V.}, \au{Ostilla-M{\'o}nico, R.}, \au{Verzicco, R.} \&
  \au{Lohse, D.}} \yr{2016}  \at{Drag reduction in numerical two-phase
  {T}aylor--{C}ouette turbulence using an {E}uler--{L}agrange approach}.
  \jt{J. Fluid Mech.}  \bvol{798},  \pg{411--435}.

\bibitem[Tisserand {\em et~al.\/}(2011)Tisserand, Creyssels, Gasteuil, Pabiou,
  Gibert, Castaing \& Chilla]{tisserand2011comparison}
{\sc \au{Tisserand, J.-C.}, \au{Creyssels, M.}, \au{Gasteuil, Y.}, \au{Pabiou,
  H.}, \au{Gibert, M.}, \au{Castaing, B.} \& \au{Chilla, F.}} \yr{2011}
  \at{Comparison between rough and smooth plates within the same
  {R}ayleigh--{B}{\'e}nard cell}.  \jt{Phys. Fluids}  \bvol{23}~(1),
  \pg{015105}.

\bibitem[Tokuhiro \& Lykoudis(1994)]{tokuhiro1994natural}
{\sc \au{Tokuhiro, A.~T.} \& \au{Lykoudis, P.~S.}} \yr{1994}  \at{Natural
  convection heat transfer from a vertical plate—i. {E}nhancement with gas
  injection}.  \jt{Int. J. Heat Mass Transfer}  \bvol{37}~(6),  \pg{997--1003}.

\bibitem[Van Den~Berg {\em et~al.\/}(2006)Van Den~Berg, Luther, Mazzitelli,
  Rensen, Toschi \& Lohse]{van2006turbulent}
{\sc \au{Van Den~Berg, T.~H.}, \au{Luther, S.}, \au{Mazzitelli, I.~M},
  \au{Rensen, J.~M.}, \au{Toschi, F.} \& \au{Lohse, D.}} \yr{2006}
  \at{Turbulent bubbly flow}.  \jt{J. Turb.} ~(7),  \pg{N14}.

\bibitem[Xie \& Xia(2017)]{xie2017turbulent}
{\sc \au{Xie, Y.-C.} \& \au{Xia, K.-Q.}} \yr{2017}  \at{Turbulent thermal
  convection over rough plates with varying roughness geometries}.  \jt{J.
  Fluid Mech.}  \bvol{825},  \pg{573–599}.

\bibitem[Zhong {\em et~al.\/}(2009)Zhong, Funfschilling \&
  Ahlers]{zhong2009enhanced}
{\sc \au{Zhong, J.-Q.}, \au{Funfschilling, D.} \& \au{Ahlers, G.}} \yr{2009}
  \at{Enhanced heat transport by turbulent two-phase {R}ayleigh-{B}{\'e}nard
  convection}.  \jt{Phys. Rev. Lett.}  \bvol{102}~(12),  \pg{124501}.

\bibitem[Zhu {\em et~al.\/}(2017{\natexlab{{\em a\/}}})Zhu, Phillips, Spandan,
  Donners, Ruetsch, Romero, Ostilla-M{\'o}nico, Yang, Lohse, Verzicco, Fatica
  \& Stevens]{zhu17afid}
{\sc \au{Zhu, X.}, \au{Phillips, E.}, \au{Spandan, V.}, \au{Donners, J.},
  \au{Ruetsch, G.}, \au{Romero, J.}, \au{Ostilla-M{\'o}nico, R.}, \au{Yang,
  Y.}, \au{Lohse, D.}, \au{Verzicco, R.}, \au{Fatica, M.} \& \au{Stevens, R. A.
  J.~M.}} \yr{2017{\natexlab{{\em a\/}}}}  \at{{AF}i{D}-{GPU}: a versatile
  {Navier-Stokes} solver for wall-bounded turbulent flows on {GPU} clusters}.
  \jt{arXiv:1705.01423} .

\bibitem[Zhu {\em et~al.\/}(2017{\natexlab{{\em b\/}}})Zhu, Stevens, Verzicco
  \& Lohse]{zhu2017roughness}
{\sc \au{Zhu, X.}, \au{Stevens, R. J. A.~M.}, \au{Verzicco, R.} \& \au{Lohse,
  D.}} \yr{2017{\natexlab{{\em b\/}}}}  \at{Roughness--facilitated local 1/2
  scaling does not imply the onset of the ultimate regime of thermal
  convection}.  \jt{arXiv preprint arXiv:1704.05126} .

\end{thebibliography}

\end{document}